\documentclass[twocolumn,twocolappendix]{aastex6}

\usepackage{amssymb,amsmath,graphicx,natbib,hyperref}
\usepackage{color}
\usepackage{enumerate}
\usepackage{booktabs, threeparttable}

\newcommand{\hii}         {\mbox{\rm H{\small II}}}

\newcommand{\Sb}          {\mbox{$\Sigma_{\rm b}$}\,}

\newcommand{\Sgas}        {\mbox{$\Sigma_{\rm gas}$}\,}
\newcommand{\Sstar}       {\mbox{$\Sigma_*$}}
\newcommand{\Ssfr}        {\mbox{$\Sigma_{\rm SFR}$}}
\newcommand{\SsfrHa}        {\mbox{$\Sigma_{\rm SFR,H\alpha}$}}
\newcommand{\SsfrSSP}        {\mbox{$\Sigma_{\rm SFR,SSP}$}}

\newcommand{\ha}          {\mbox{H$\alpha$}}
\newcommand{\hb}          {\mbox{H$\beta$}}
\newcommand{\msunperpcsq} {\mbox{\rm M$_\odot$~pc$^{-2}$}}
\newcommand{\msunperkpcsq} {\mbox{\rm M$_\odot$~kpc$^{-2}$}}

\newcommand{\av}          {\mbox{${\rm A_V}$\,}}
\newcommand{\Smol}        {\mbox{$\Sigma_{\rm mol}$}\,}
\newcommand{\SmolAv}        {\mbox{$\Sigma_{\rm mol,Av}$}\,}

\begin{document}

\title{SDSS-IV MaNGA: A star formation - baryonic mass relation at kpc scales. }

\author{J. K. Barrera-Ballesteros\altaffilmark{1}, T. Heckman\altaffilmark{2}, S. F. S\'anchez\altaffilmark{1}, N. Drory\altaffilmark{3}, I. Cruz-Gonzalez \altaffilmark{1}, L. Carigi\altaffilmark{1},  R. A. Riffel\altaffilmark{4,5}, M. Boquien\altaffilmark{6}, P. Tissera\altaffilmark{7,8}, D. Bizyaev\altaffilmark{9}, Y. Rong\altaffilmark{10}, N. F. Boardman\altaffilmark{11}, P. Alvarez Hurtado\altaffilmark{1}  \& the MaNGA team}

\altaffiliation{$^{1}$ Instituto de Astronom\'ia, Universidad Nacional Aut\'onoma de M\'exico, A.P. 70-264, 04510 M\'exico, D.F., M\'exico}
\altaffiliation{$^{2}$Department of Physics \& Astronomy, Johns Hopkins University, Bloomberg Center, 3400 N. Charles St., Baltimore, MD 21218, USA}
\altaffiliation{$^{3}$University of Texas at Austin,McDonald Observatory, 1 University Station, Austin, TX 78712, USA}
\altaffiliation{$^{4}$Departamento de F\'isica, CCNE, Universidade Federal de Santa Maria, 97105-900, Santa Maria, RS, Brazil}

\altaffiliation{$^{5}$Laborat\'orio Interinstitucional de e-Astronomia - LIneA, Rua Gal. Jos\'e Cristino 77, Rio de Janeiro, RJ - 20921-400, Brazil}
\altaffiliation{$^{6}$Universidad de Antofagasta, Centro de Astronom\'ia, Avenida Angamos 601, Antofagasta 1270300, Chile}
\altaffiliation{$^{7}$Instituto de Astrof\'isica, Pontificia Universidad Cat\'olica de Chile, Santiago, Chile}
\altaffiliation{$^{8}$Centro de Astro-Ingenier\'ia, Pontificia Universidad Cat\'olica de Chile, Santiago, Chile}
\altaffiliation{$^{9}$Apache Point Observatory, P.O. Box 59, Sunspot, NM 88349}
\altaffiliation{$^{10}$Chinese Academy of Sciences, National Astronomical Observatories of China, 20A Datun Road, Chaoyang District, Beijing 100012, China}
\altaffiliation{$^{11}$University of Utah, Department of Physics and Astronomy, 115 S. 1400 E., Salt Lake City, UT 84112, USA}
\email{jkbarrerab@astro.unam.mx}
\begin{abstract}
Star formation rate density, \Ssfr, has shown a remarkable correlation with both components of the baryonic mass at kpc scales (i.e., the stellar mass density, and the molecular gas mass density; \Sstar, and \Smol, respectively) for galaxies in the nearby Universe. In this study we propose an empirical relation between \Ssfr\, and the baryonic mass surface density (\Sb =\SmolAv + \Sstar; where \SmolAv is the molecular gas -- derived from the optical extinction, \av) at kpc scales using the spatially-resolved properties of the MaNGA survey -- the largest sample of galaxies observed via Integral Field Spectroscopy (IFS, $\sim$ 8400 objects). We find that \Ssfr\, tightly correlates with \Sb. Furthermore, we derive an empirical relation between the \Ssfr\, and a second degree polynomial of \Sb yielding a one-to-one relation between these two observables. Both, \Sb and its polynomial form show a stronger correlation and smaller scatter with respect to \Ssfr\,than the relations derived using the individual components of \Sb. Our results suggest that indeed these three parameters are physically correlated, suggesting an scenario in which the two components of the baryonic mass regulate the star-formation activity  at kpc scales. 

\end{abstract}

\section{Introduction}
\label{sec:Intro}

Understanding what are the physical scenarios that describe the star formation activity in galaxies is fundamental to explain their evolution throughout their lifetime. In turn, depending on the explored spatial scale, there have been mainly two different, yet complementing, scenarios that explain the star formation in galaxies. Broadly speaking, one possibility is that the amount of newly formed stars in a galaxy is set primarily by the local amount of cold gas available to create that newly-born population. On the other hand, star formation can also be affected by global properties, for instance the dynamical structure of the disk of the galaxy as a whole \citep[][and references therein]{Kennicutt_2012}. A complementary approach suggests that star-formation is self-regulated \citep[e.g.,][]{Dopita_1985,Dopita_1994,Silk_1997,Ostriker_2010}. In this scenario, momentum injection from massive stars balances the hydrostatic pressure due to the disk's weight.  

Observationaly, empirical scaling relations have been fundamental to explore the role of the baryonic mass in the physical processes that yield the amount of newly-born stars in the Universe. The Schmidt-Kennicutt (SK) relation is the most well-known of those relations. It provides a strong correlation between the observed star formation rate (SFR) and the amount of cold gas \citep{Schmidt_1959,Kennicutt_1998}. When using intensive measurements -- properties average across a certain area projected in the sky -- (SFR surface density, \Ssfr; and gas mass surface density, \Sgas), the SK relation follows a similar slope across several orders of magnitude including a wide range of galaxy morphologies and types for normal star-forming galaxies \citep{Gao_2004}. It can vary when exploring extreme starburst galaxies or galaxies at high redshift \citep[e.g.,][]{Daddi_2010,Genzel_2010}. Thanks to radio-interferometric surveys, there is also a wealth of data indicating that the SK relation is also valid at kpc scales for a large sample of extragalactic star forming sources \citep[e.g., HERACLES, THINGS, and EDGE-CALIFA surveys][]{Leroy_2008, Walter_2008, Bolatto_2017}.  At a very first order, the SK law and its counterpart at kpc scales, the resolved SK relation (rSK) can be describe by the first scenario described above.   

Another star-forming scaling relation is the one that shows a tight correlation between the galaxy-integrated SFR and the total stellar mass ($\mathrm{M_{\ast}}$). The so-called star formation main sequence (SFMS) has been derived for thousands of galaxies included in the DR7 Sloan Digital Sky Survey \citep[SDSS, e.g., ][]{Kauffmann_2003, Brinchmann_2004}. Note that in the SFR-$\mathrm{M_{\ast}}$ plane it is also observed a cloud of massive galaxies with little SFR (known as the retired sequence of galaxies). In the last decade, thanks to Integral Field Unit (IFU) observations in large samples of galaxies, it has been possible to determine the existence of a local counterpart of the SFMS: the resolved SFMS (rSFMS, \Ssfr\, vs \Sstar) for a large sample of galaxies \citep{Sanchez_2013, Wuyts_2013}. The existence of the rSFMS -- as well as the spatially-resolved version of the retired sequence -- may also indicate a more intimate correlation between the current star formation and the star formation history of a galaxy at kpc scales \citep[e.g., ][]{Cano-Diaz_2016}. Recently, with large IFU datasets and the combination with direct observations of spatially resolved  molecular gas, different studies have established that \Ssfr\, is mainly correlated to \Sstar\, than \Sgas \citep[e.g., ][]{Dey_2019,Bluck_2020}. Similar studies have also noted the significant relation at kpc scales between the molecular mass gas density and the stellar mass density \citep[e.g.,][]{Lin_2019, BB2020}. 

The above two star-forming relations are not the only ones that correlate at kpc scales the \Ssfr\, with the baryonic mass component in galaxies. Different authors have explored the relation between the \Ssfr\, and the \Sgas\,\Sstar\, product or similar relations with the baryonic mass \citep[e.g., ][]{Matteucci_1989,Shi_2011, Shi_2018}. Since the combination of the baryonic densities scales with the disk hydrostatic mid-plane pressure, these studies were aimed to explore whether star-formation follows a self-regulated scenario. In this scenario, the star-formation is locally regulated by the interplay between the mid-plane pressure produced by the baryonic mass and the momentum flux due to supernovae explosions from the most massive stars \citep{Ostriker_2010,Shetty_2012,Kim_2013}.       

Despite the great advance in our understanding of the star-formation from these works, there has been a lack of systematic studies exploring the relation between the baryonic mass (\Sb = \mbox{\Sstar + \Sgas}) and \Ssfr\, at kpc scales. Such a relation can shed some light regarding the most likely physical scenario of star formation at kpc scales. In this study, we take advantage of the SDSS-IV MaNGA survey \citep{Wake_2016}, the largest IFU dataset to date to explore this correlation. Furthermore we also investigate the impact of a correlation of the \Sgas with a non-linear combination of \Sb. This article is structured as follows. In Sec.~\ref{sec:Sample_Data} we present our sample selection as well as an overview on the main features of the datacube for each galaxy included in the MaNGA survey. In Sec.~\ref{sec:AR} we show the observables derived from the datacubes required for this study and our main results, while in Sec.~\ref{sec:Discussion} we discuss our main findings. The main conclusions of this article are presented in Sec.~\ref{sec:Conclusion}.

\section{Sample and datacubes}
\label{sec:Sample_Data}

The galaxies selected for this study are drawn from the latest sample of targets observed in the MaNGA survey \citep{Bundy_2015} which is part of the fourth generation of the Sloan Digital Sky Survey \citep[SDSS IV, ][]{Blanton_2017}. This sample includes galaxies observed from  March of 2014 to September of 2019 (8405 datacubes). This sample corresponds to the internal release within the collaboration known also as MaNGA product launch (MPL-9). The MaNGA survey has been designed to obtain IFS observations for more than 10,000 galaxies. A detailed description of the selection criteria for this survey is presented in \citep{Wake_2017}.

\begin{figure*}
 \begin{minipage}[c]{0.48\textwidth}
\includegraphics[width=\linewidth]{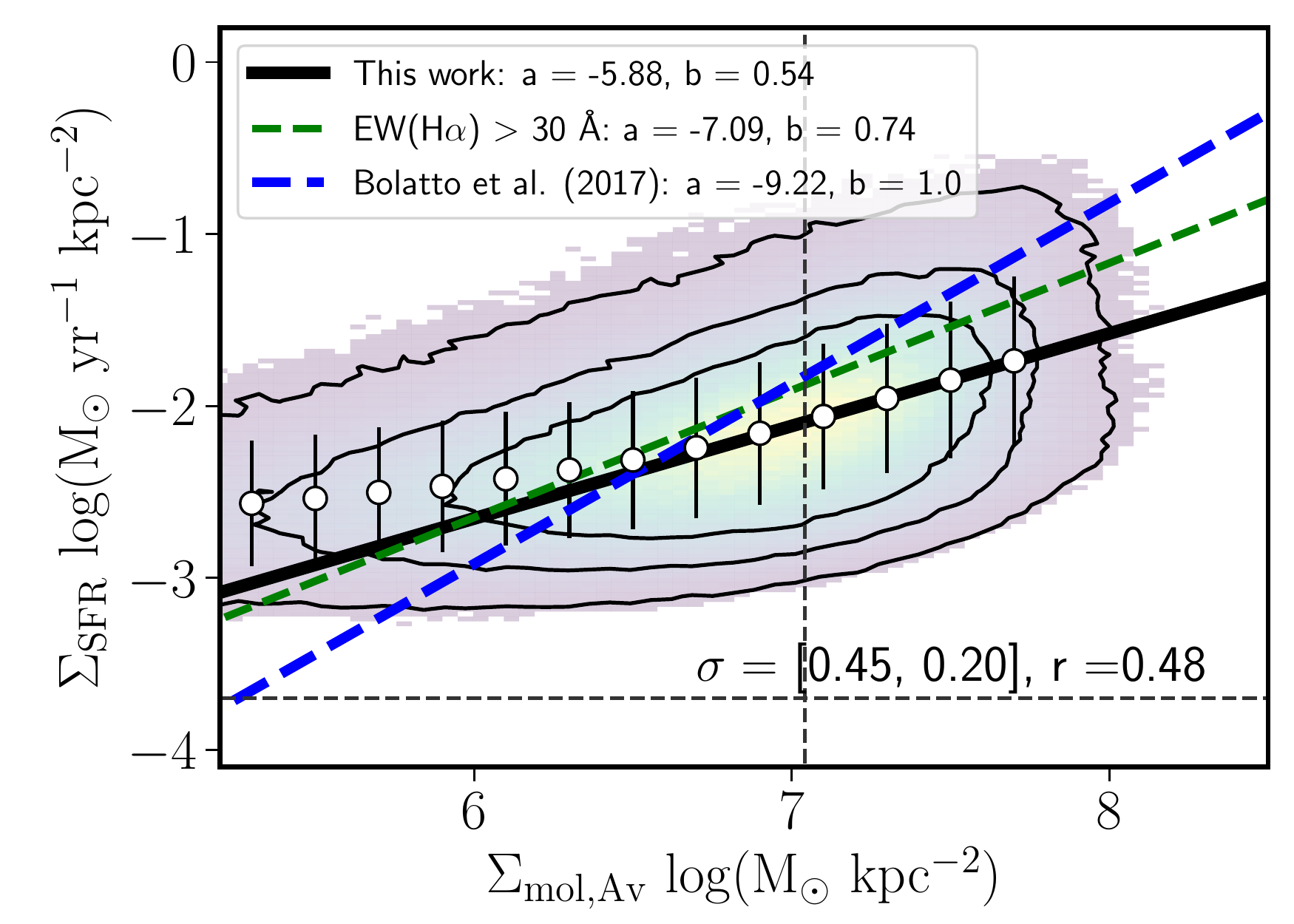}
\end{minipage}
 \begin{minipage}[c]{0.48\textwidth}
\includegraphics[width=\linewidth]{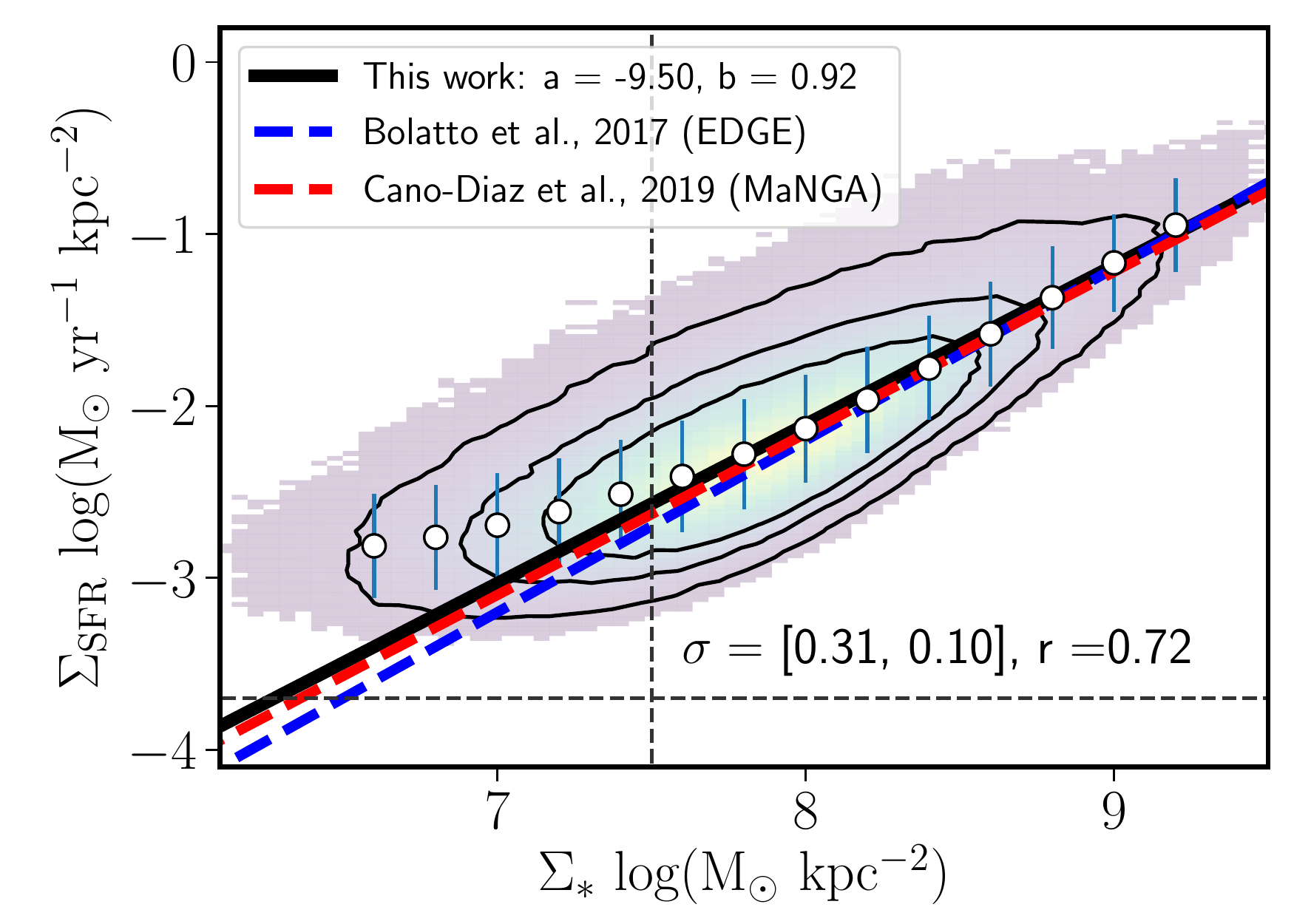}
\end{minipage}
\caption{Distributions of \Ssfr\, versus \Sgas (rSK, left panel) and versus \Sstar (rSFMS, right panel) for more than 1.1$\times$10$^6$ spaxels, selected from 2640 galaxies included in the MaNGA survey. In both panels the distributions are color coded according to the spaxels density. Black contours enclose 95\%, 80\%, and 60\% of the distributions respectively. The white circles with error bars represent the median \Ssfr\, for bins of \Sgas and \Sstar, respectively. The black lines represent the best fit derived from the medians (with circles). These fits are derived from bins with \Sgas and \Sstar\ larger than the gray vertical dashed lines (see Sec.\ref{sec:single}). The slopes from both fits are in good agreement with those derived using direct observations of molecular gas in smaller samples of galaxies \citep[][dashed red and blue lines, respectively]{Bolatto_2017,Cano-Diaz_2019}.
}
\label{fig:sKS_SFMS}    
\end{figure*}

The observations of the MaNGA survey are taking place at the Apache Point Observatory using its 2.5-m telescope \citep{2006AJ....131.2332G}. A detailed description of the instrumentation of the survey can be found in \cite{2015AJ....149...77D}. For a detailed explanation of the data strategy the reader is refereed to \cite{2016AJ....152...83L}.  The MaNGA reduction pipeline includes wavelength calibration, corrections from fiber-to-fiber transmission, subtraction of the sky spectrum and flux calibration \citep{2016AJ....151....8Y}. The final product is a datacube with $x$ and $y$ coordinates corresponding to the sky coordinates and the $z$-axis corresponds to the wavelength. Its final spaxel size is 0.5\arcsec with a spatial resolution of 2.5\arcsec corresponding to a mean physical scale of $\sim$ 2 kpc.
\section{Analysis and Results}
\label{sec:AR}
\subsection{Spatially-resolved and integrated properties from the observed galaxies}
\label{sec:Data}
The local observables for this study are determined by applying the \textsc{PIPE3D} analysis pipeline to the MaNGA datacubes. A full description of this pipeline, its stellar and emission line fitting, uncertainties estimations, and derivation of physical properties can be found in \cite{Sanchez_2016_PIPE3D}. An overview on how the two-dimensional distributions of stellar and gas component are derived for MaNGA data can be found in \cite{BB2016, BB_2018}. In particular for this study, we use the stellar surface mass density (\Sstar), star formation rate density (\Ssfr(SSP)), and the specific SFR (sSFR(SSP)) derived from the single-population stellar (SSP) analysis. \Ssfr(SSP) is obtained as the ratio between the mass of stars formed in the last temporal bin (0.06 Gyr) and the total mass of stars formed across the cosmic time. For this SSP analysis the pipeline adopts a Salpeter IMF. From the emission lines analysis we obtain for each spaxel the extinction-corrected \Ssfr\, derived from the \ha\, emission line luminosity, \Ssfr(\ha), and the \ha\, equivalent width, EW(\ha). We follow \cite{Kennicutt_1998_SFR} to derive \Ssfr, where a Salpeter IMF is also adopted. In \cite{BB2020} we derived a linear correlation between \Sgas (\mbox{$\Sgas =\mathrm{\Smol + \Sigma_{H_I}}$}) and the optical extinction (\av). \av was derived from the Balmer decrement, whereas $\mathrm{\Sigma_{H_2}}$ was derived from CO resolved maps observed within the EDGE survey \citep{Bolatto_2017}. We found that \Sgas (\msunperpcsq) $\sim$ 26 \av (mag) for radial scales.  With the same dataset, we derived in Barrrera-Ballesteros et al. (in prep.) a linear log-log calibrator for the molecular gas mass density with \mbox{$\SmolAv (\msunperpcsq) = 10^{1.37} [\av\,(\mathrm{mag})]^{2.3}$} using a least trimmed square (LTS) fitting at kpc scales \citep[]{Cappellari_2013}. In comparison to the linear calibrator, it improves the estimation of \Smol for high values of \av. Even more, we found that this calibration depends on the inclination therefore we use it only for low-inclined galaxies.  All the intensive observables are corrected by inclination following \cite{BB2016}.

For this study we select those regions (spaxels) that we consider {\it bona fide} star-forming in low-inclined galaxies ($b/a <$ 0.45\footnote{$b/a$ for each MaNGA galaxy is obtained from the extended NASA-Sloan catalog  \citep[NSA][]{Blanton_2005}: http://www.nsatlas.org}). A spaxel is consider as star-forming if: {\it(i)} the \mbox{\rm [N{\small II}] / \ha} and \mbox{\rm [O{\small III}] / H$\beta$} emission line flux ratios are below the \cite{Kewley_2001_BPT} demarcation line in a BPT diagram \citep{BPT_1981}; {\it(ii)} the \ha\, and H$\beta$ emission lines have a signal-to-noise ratio larger than 3; {\it(iii)} \mbox{\ha/H$\beta$ $>$ 2.86}; and {\it(iv)} an EW(\ha) $>$ 14 \AA. Based on the relation derived in \cite{Sanchez_2013}, this selection criteria is equivalent to select regions with a specific SFR, sSFR $\gtrsim$ 10$^{-10.47}$ \mbox{yr$^{-1}$}. These criteria lead to select more than \mbox{$1.1 \times\, 10^6$} star-forming spaxels located in 2640 galaxies included in the MaNGA sample. We should note that due to the spatial resolution of this survey ($\sim$ 2.5\arcsec), the number of spaxels overestimates the number of independent \hii\, regions detected in the survey by at least a factor $\sim$ 5.

\subsection{Single-variable scaling relation: rSK and rSFMS}
\label{sec:single}

In Fig.~\ref{fig:sKS_SFMS} we show the rSK and the rSFMS derived for our sample of galaxies (left and right panels, respectively). The outermost black contour encloses 95\% of the distributions. The solid circles represent the median \Ssfr\, for bins of equal width (0.2 dex) on \SmolAv and \Sstar, respectively. The black solid lines represent a least-square linear fit of those median values in log scales. We use the relation \mbox{$\log(\Ssfr) = a +b \log(\Sigma)$}, where $\Sigma$ represents both, \Smol\, and \Sstar, respectively.  Following \cite{Cano-Diaz_2019}, in order to perform the above fit for both relations we use those bins where we consider that we have reliable estimations of \SmolAv and \Sstar. For the rSK relation we select those bins with \mbox{$\SmolAv > 10^{7.0} \msunperkpcsq$}. This limit is motivated by the gas calibrator used in this study. In Barrera-Ballesteros et al. (in prep.) we show that due to the sensitivity of the CO observations \citep[from the EDGE survey][]{Bolatto_2017}, the calibrator may not be reliable at \SmolAv smaller than this limit. On the other hand, to perform the fitting in the rSFMS \cite{Cano-Diaz_2019} used bins with \mbox{$\Sstar \,>\, 10^{7.5} \msunperkpcsq$}. They noted a non-physical driven flattening of the rSFMS below the mentioned \Sstar\ due to detection limits in each of the observables. Both of these limits are represented as vertical dashed lines in Fig.\ref{fig:sKS_SFMS}. In both panels, the horizontal dashed line represent the of \ha\, surface brightness detection limit for the MaNGA survey \citep[$\sim 10^{37}\mathrm{erg\,s^{-1}\,kpc^{-2}}$,][]{Cano-Diaz_2019}. Finally, we note that the selection criteria based on the EW(\ha) naturally results in tight relation since we are considering only regions above a certain sSFR.

Both scaling relations present significant Pearson correlation coefficients ($r$). In fact, our rSFMS shows a higher $r$ coefficient in comparison to previous studies using MaNGA data \citep[e.g., ][]{Cano-Diaz_2019, Lin_2019}. However, the $r$ coefficient for the rSK relation is smaller in comparison to the value derived for the rSFMS relation (0.48 vs 0.72). This could be caused by the larger scatter observed in the rSK compared to the rSFMS. We also note that in comparison to other relations presented in the literature our rSK shows a smaller correlation factor \citep[e.g.,][]{Lin_2019}. We compare the best fit for each relation with those that illustrate the trends observed from the EDGE survey \citep{Bolatto_2017} which makes use of CO maps to estimate \Sgas in a sample of 123 galaxies. The best fit for the rSK derived using the MaNGA data is flatter in comparison to the relation derived by \cite{Bolatto_2017}. We suggest that this flattening is due to the nature of our dataset. For low values of \Ssfr\, we could be sampling physical regions for which \ha\, flux emission could be potentially polluted from processes other than young stellar emission \citep[e.g., diffuse \ha\, emission from old stars,][]{Lacerda_2018}. In turn, this causes an underestimation of \Sgas from the calibrator. By diluting \hii\, regions along with diffuse \ha\ emission, the spatial resolution can also have an impact in the scatter of the rSK \citep[e.g., ][]{Vale-Asari_2020}. However, for large values of \Ssfr, \av recovers the expected value of \Smol since at high \Ssfr\, the amount of dust traces \Smol. Indeed, when selecting spaxels with EW(\ha) larger than 30 \AA\, the best linear fit is steeper to the one derived from the median values (green dashed line) and is close to the linear relation reported using direct CO measurements by the EDGE survey \citep{Bolatto_2017}. On the other hand, the rSFMS is in very good agreement with those derived previously using the EDGE and MaNGA datasets \citep[][]{Bolatto_2017, Cano-Diaz_2019}.       

In both panels we also annotate the scatter of the residuals, $\sigma$. We use two measurements, the standard deviation and the variance. By comparing $\sigma$ between these two relations we note that the rSK shows a larger scatter in comparison to the rSFMS ([0.45, 0.20] vs [0.31, 0.10]). 

The scatter of the rSK is only $\sim$0.01 dex larger in comparison to the rSK relation derived using direct observations of molecular gas in a smaller sample of MaNGA galaxies \citep{Lin_2019}. In other words, the scatter of the relation remains similar as more galaxies are probed. On the other hand, it may be the case that the gas estimator could artificially increase the scatter. However, it would very difficult to provide direct spatially-resolved observations of the gas density for the sample of galaxies presented in this study. So far, the largest overlap of IFS data with spatial resolved CO observations comprises only 126 galaxies, from the EDGE-CALIFA survey \citep{Bolatto_2017}. Other similar efforts, like the ALMAQUEST compilation, comprises 47 MaNGA galaxies extracted from different scientific projects \citep{Lin_2019}.  

We also note that the scatter of the rSFMS is reduced in comparison to other estimations of the rSFMS using the MaNGA dataset \citep[e.g.,][]{Hsieh_2017, Lin_2019, Cano-Diaz_2019}. This suggests that using a larger sample of galaxies can only mildly reduce the statistical scatter of this relation. 

\subsection{The linear \Ssfr-\Sb relation}
\label{sec:Sb}
\begin{figure}
\includegraphics[width=\linewidth]{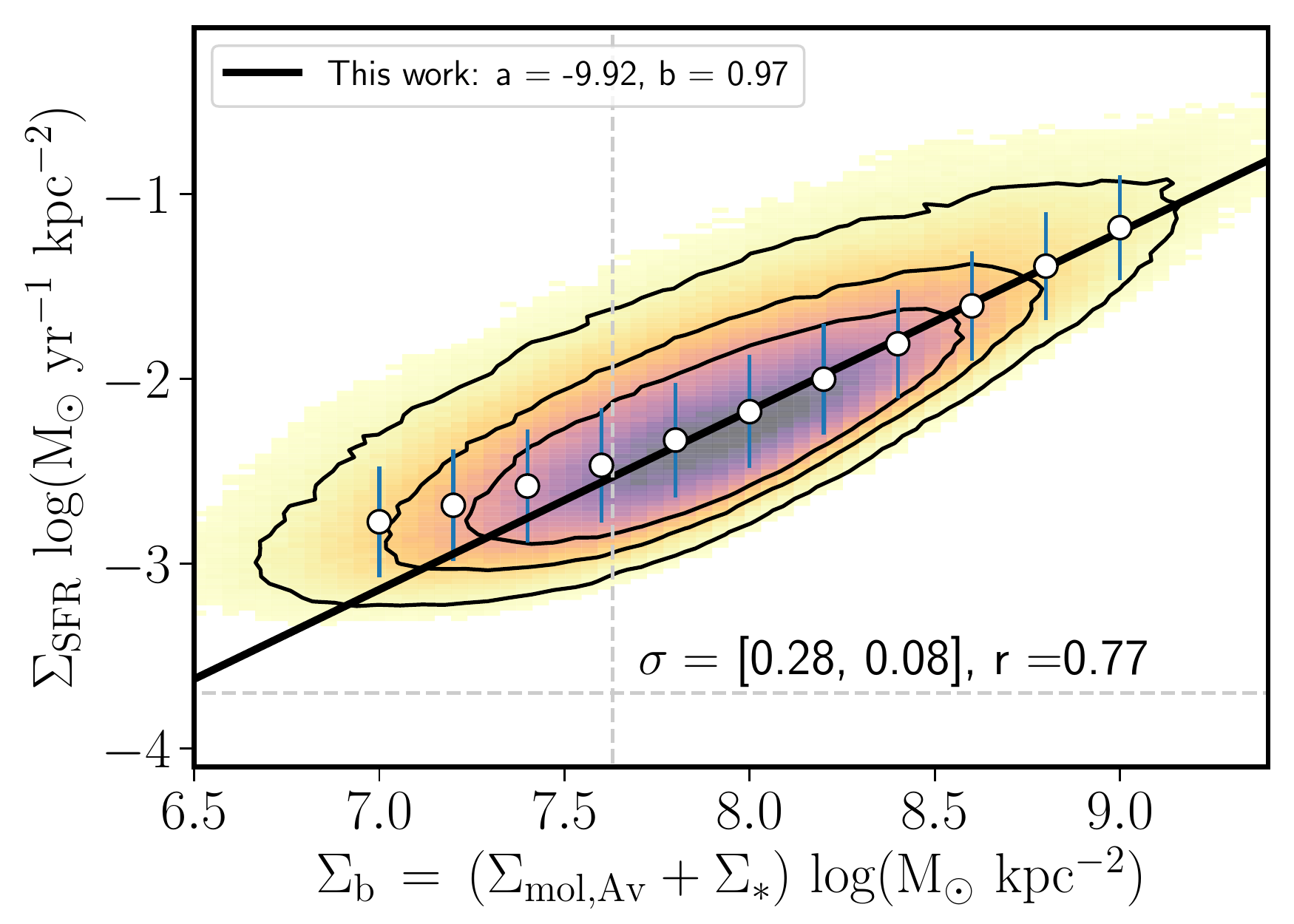}
\caption{The \mbox{\Ssfr - \Sb} relation for the sample of spaxels included in this study. The description of the lines and symbols is similar as the one presented for panels in Fig.\ref{fig:sKS_SFMS}. In comparison to the single-variable scaling relations, \Sb shows a higher correlation factor as well as a tighter relation with \Ssfr.}
\label{fig:SbSF}    
\end{figure}

One of the goals of this study is to probe what is the best relation of the baryonic mass (\mbox{\Sb = \SmolAv + \Sstar}) with respect to the \Ssfr. First, we show the direct relation between these two parameters in Fig.~\ref{fig:SbSF}. To our knowledge, this is the first time that this scaling relation is explored for such a large sample of galaxies at kpc scales. The distribution of this relation is drawn from the same regions used in the relations explored in Fig.\ref{fig:sKS_SFMS}. We note that the distribution of spaxels in the \mbox{\Ssfr - \Sb} plane does not show the strong flattening observed at low values of \Sb, in the rKS and rSFMS relations (see Fig.~\ref{fig:sKS_SFMS}). 

The distribution seems to follow a tighter relation than the single-parameter local star-forming relations explored in Fig.~\ref{fig:sKS_SFMS} (i.e., lower scatter). In fact, the $r$-coefficient indicates a stronger correlation between \Ssfr\, and \Sb than the relations derived in Sec.\ref{sec:single}. For instance, the  $r$-coefficient is larger for this relation than for the rSFMS (0.77 vs 0.72). We follow a similar procedure to derive the best fit relation as in Sec.~\ref{sec:single}. We choose a \Sb threshold to select the bins to perform the fit (\mbox{\Sb $\gtrsim 10^{7.6}$ \msunperkpcsq}, this is the sum of the thresholds in Sec.~\ref{sec:single}). However, we should note that the results are not strongly affected by the implementation of this limit in the analysis. The best fit parameters (slope and zero-point) are shown in Fig.~\ref{fig:SbSF}. Both parameters of the best fit are similar to those derived from the rSFMS (see Fig.~\ref{fig:sKS_SFMS}). This is expected since the main driver in the baryonic mass density is \Sstar. Nevertheless, it is important to note the role of \Sgas\, in shaping the \mbox{\Ssfr\, - \Sb} relation at low values of \Sb as well as increasing the slope of the best relation making it close to one. This can also be noted in the decrease in the scatter of the \Sb-\Ssfr\, relation in comparison to the rSFMS. Finally, the scatter of the residuals with respect to this fit is smaller than the rSFMS ($\sigma = [0.28, 0.08]$ vs $\sigma = [0.31, 0.10]$) indicating a tighter relation than the ones derived by individual components of \Sb.

\subsection{Impact of a quadratic term in the \mbox{\Sb - \Ssfr} relation}
\label{sec:xnew}

As we mentioned in Sec.~\ref{sec:Intro}, there are different studies exploring scaling relations between the star formation and the \Sgas\Sstar\, product or similar relations. These relations have been studied in order to explore the self-regulated star-formation scenario. In this section, we explore a rather generic approach. We investigate the relation of \Ssfr\, with a second degree polynomial of \Sb. Using this functional form, we explore both the dependence of \Ssfr\, with respect of mixed quadratic terms (\Sgas\Sstar) as well as quadratic terms of \Sgas$^2$ and \Sstar$^2$. 

\begin{figure}
\includegraphics[width=\linewidth]{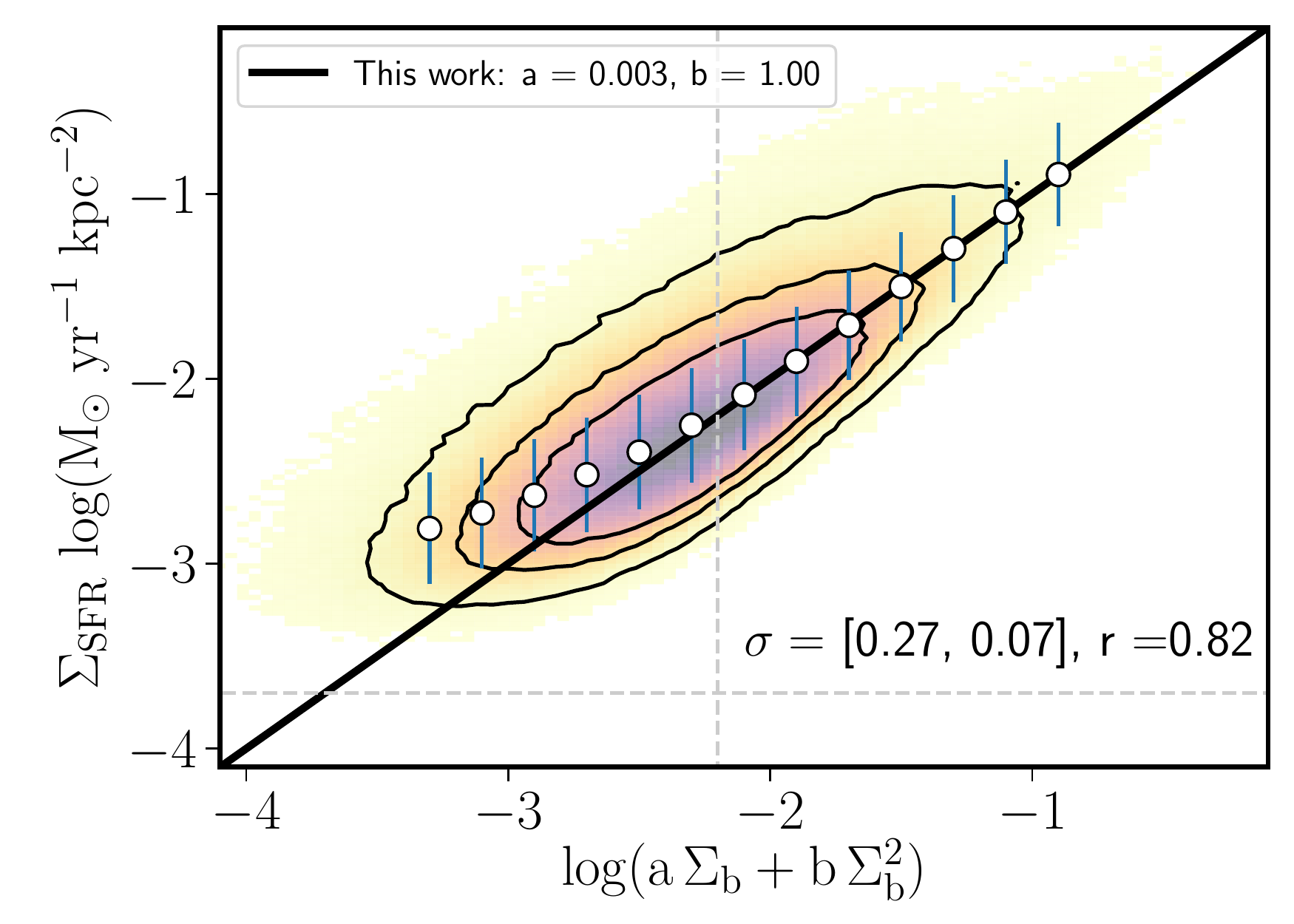}
\caption{A general relation between the \Ssfr\, and \Sb. The x-axis ($\mathrm{x_{new}}$) represents the best fit of a 2nd degree polynomial of \Sb for the observed \Ssfr\, which yields a one-to-one relation. As in previous plots, the contours enclose 95\%, 80\%, and 60\% of the distribution. The white circles show the median \Ssfr\, for bins of $\mathrm{x_{new}}$. The black line shows the best fit for these median values. The description of the lines and symbols is similar as the one presented for panels in Fig.\ref{fig:sKS_SFMS}. The scatter and $r$ coefficient of this one-to-one relation resambles those valued derived for the \Ssfr-\Sb relation.}  
\label{fig:SbSFfit}    
\end{figure}

In Fig.~\ref{fig:SbSFfit} we plot the relation between the x-axis defined as \mbox{$\mathrm{x_{new} = a\,\Sb + b\,\Sb^{2}}$} against \Ssfr. The fit between \mbox{$\mathrm{x_{new}}$} and \Ssfr\, is obtained by using the median values of \Sb and \Ssfr\, derived in Fig.~\ref{fig:SbSF}. This fit yields the values of \mbox{$\log(a) = -10.21$} and \mbox{$\log(b) = -20.37$}. The spaxels in the \mbox{$\Ssfr-\mathrm{x_{new}}$} plane lies in a well-defined linear trend. In fact, the best linear fit derived from the median \Ssfr\, for different \mbox{$\mathrm{x_{new}}$} bins yield a one-to-one relation (black solid-line in Fig.~\ref{fig:SbSFfit}). In contrast to the single-variable scaling relations (rKS, and rSFMS; see Sec.~\ref{sec:single}), the above relation does not show strong deviations or flattening at low values of  \mbox{$\mathrm{x_{new}}$}. We perform the same analysis using each of the components of \Sb separately. We find that by fitting a quadratic polynomial using \Sstar\, or \SmolAv, as independent variables, there is no significant reduction in the scatter and neither it is possible to obtain a one-to-one relation from the medians as we find using \Sb. Therefore, our results indicate that a better representation between the \Ssfr\, and the baryonic mass is given by a second degree polynomial rather than by considering each of its components separately. 

The best-fit relation between \mbox{$\Ssfr-\mathrm{x_{new}}$} yields a similar scatter and a larger $r$-correlation factor compare to those derived with \Sb alone. To test how this relation can be affected by the number of sampled galaxies, we derive the same figures in this analysis using the dataset from the previous internal release (MPL8, $\sim$ 6400 galaxies). We find the same results in terms of slopes and scatters as those drawn from the current sample. These results may suggest that the residuals observed in the \Ssfr-\Sb relation as well as the \mbox{$\Ssfr-\mathrm{x_{new}}$} relation may be due to the statistical distribution of the observables. We note that despite the linear slope derived including a quadratic term, the best-fit coefficients are small, in particular the factor that multiplies the quadratic term of \Sb is several orders of magnitude smaller than the factor that accompanies the linear term of \Sb. The derivation of a slope of one is expected since, as we mention in Sec.~\ref{sec:Sb}, the stellar mass density is the dominant term in \Sb, in particular for regions with large star formation rates.

\begin{figure*}
 \begin{minipage}[c]{0.48\textwidth}
\includegraphics[width=\linewidth]{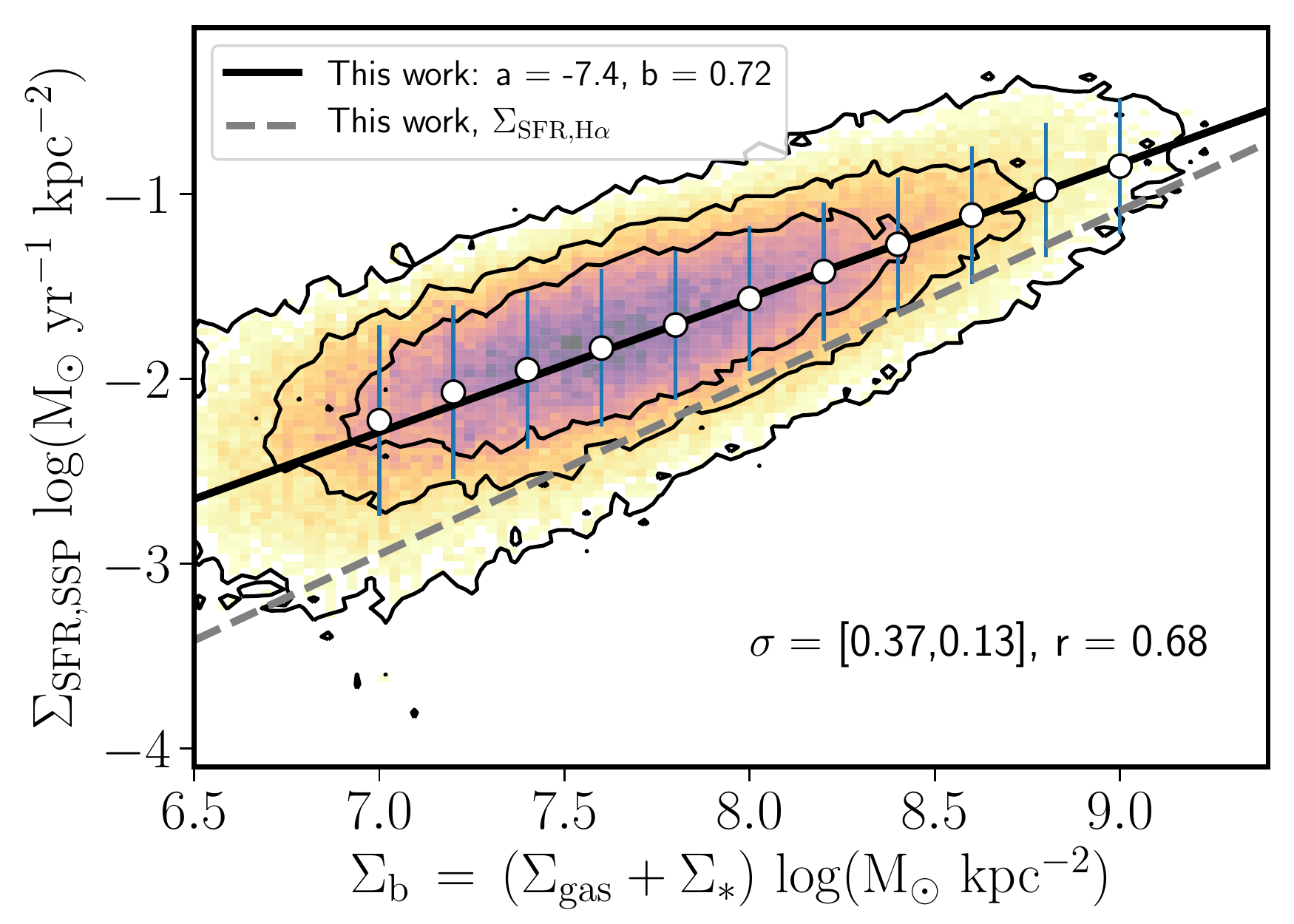}
\end{minipage}
\begin{minipage}[c]{0.49\textwidth}
\includegraphics[width=\linewidth]{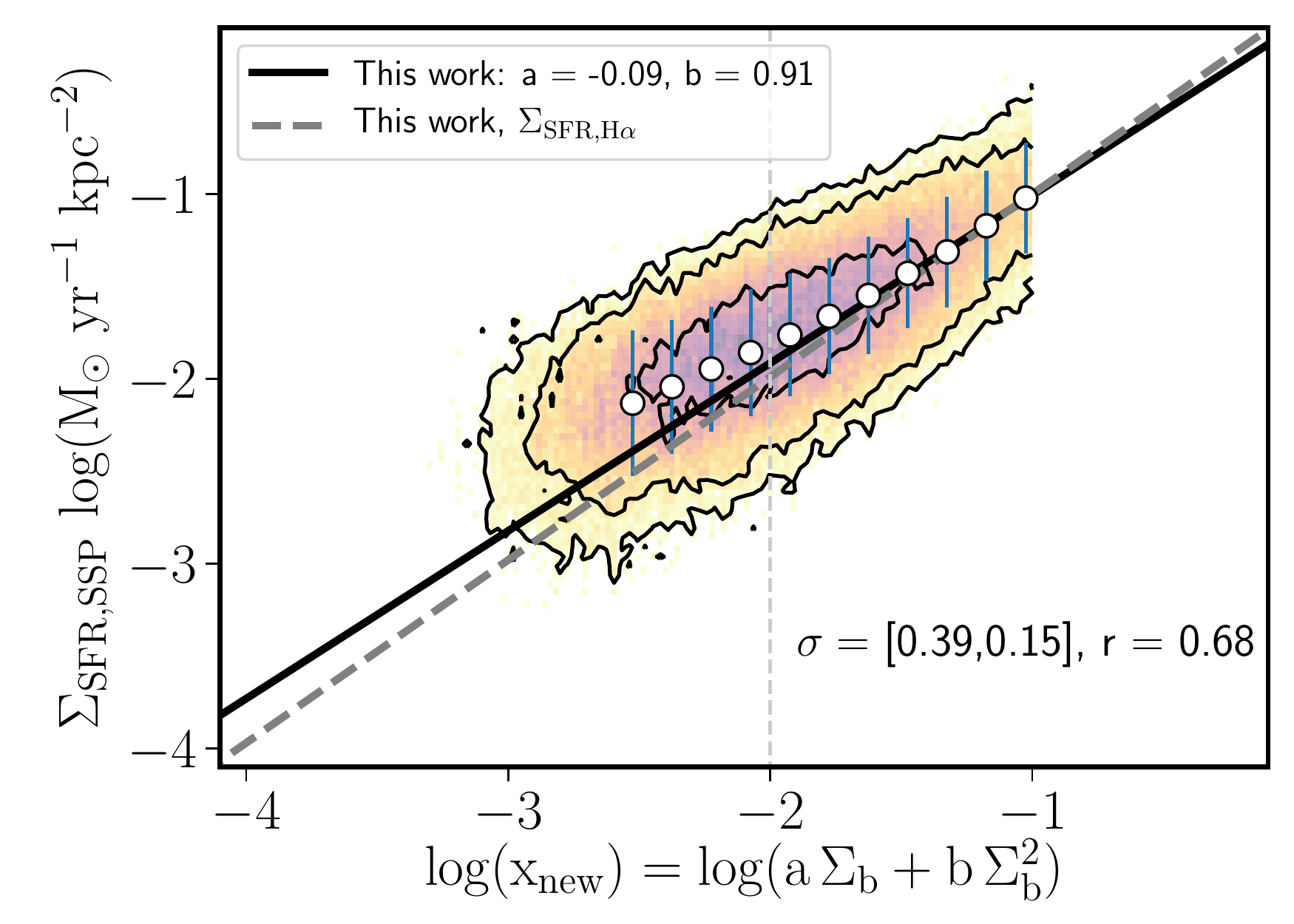}
\end{minipage}
\caption{The star-forming scaling relation derived in this study using \SsfrSSP\, instead of \SsfrHa. ({\it left panel}) The \Ssfr-\Sb relation. The dashed line represents the best-fit relation derived using \SsfrHa\, (see Fig.~\ref{fig:SbSF}). ({\it right panel}) The relation between \Ssfr-$\mathrm{x_{new}}$. As in the previous panel, the dashed line represents the best-fit relation as presented in Fig.~\ref{fig:SbSFfit}. Independent of the observable used to determine \Ssfr, the general relation between the baryonic mass and the star formation at kpc scales holds.   
}
\label{fig:RSsfrSSp}    
\end{figure*}

\subsection{An independent measure of \Ssfr\, via SSPs}
\label{sec:SSPs}

In the previous sections we derive the scaling relations between the \Ssfr\, and the different functions of the baryonic mass using the \ha\, luminosity as the observable to derive \Ssfr. Similarly, we estimate \Sgas\, from the \ha/\hb\ emission lines ratio (Balmer decrement). Therefore it can be the case that since we are using similar observables to determine the above scaling relations we could be inducing such relations. In order to test this, we use in this section another estimation of \Ssfr. As we mention in Sec.~\ref{sec:Data}, the \textsc{Pipe3D} data analysis pipeline allow us, through the fitting of SSPs to the stellar continuum of each spaxel, to determine, among other properties of the stellar component, the average star formation at different cosmic times \citep[this is, their star-formation histories, SFHs; ][]{IbarraMedel_2019}. For the purposes of this study we understand the \SsfrSSP\ as the fraction of the latest stellar burst measured by the SSP fitting \citep[i.e., the fraction of stars formed in a span of time smaller than $\sim$32~Myr, ][]{Gonzalez-Delgado_2016}. Besides the selection criteria describe in Sec.\ref{sec:Data}, for this section we only consider spaxels with a sSFR(SSP) $> 10^{-10} \mathrm{yr^{-1}}$. As result, the sample for this experiment consist of 3.1$\times10^{5}$ spaxels located in 2098 galaxies. In Appendix \ref{app:SFRs} we compare the estimation of \Ssfr using SSPs (\SsfrSSP) with the one derived using \ha\, luminosity (\SsfrHa). We note that they strongly correlate with each other ($r$ = 0.73). In particular, they are similar at large values. In average, \SsfrSSP\, is overestimated by a factor of 1.3  in comparison to \SsfrHa. Different studies showed that scaling relations at global and local scales derived using either \SsfrSSP\, or \SsfrHa\, are similar \citep[e.g., the (r)SFMS; see ][]{Gonzalez-Delgado_2014,Gonzalez-Delgado_2016, Sanchez_2018}.

In the left and right panels of  Fig.~\ref{fig:RSsfrSSp} we plot the same relations from Figs~\ref{fig:SbSF} and \ref{fig:SbSFfit} using \SsfrSSP, respectively. We find similar results when we derive the \Ssfr-\Sb relation using \SsfrSSP\, instead of \SsfrHa. However, the distribution is above the relation derived using \SsfrHa\, (see dashed line). Although the trend is similar, the slope of this relation is shallower than the one derived using the \ha\, proxy for \Ssfr\, it also presents a scatter larger than the one derived in Sec.~\ref{sec:Sb}. In right panel of  Fig.~\ref{fig:RSsfrSSp}  we plot the \Ssfr-$\mathrm{x_{new}}$ relation following the same procedure as in Sec.~\ref{sec:xnew}. We note that coefficients of the fit for $\mathrm{x_{new}}$ are different as those derived in Sec.\ref{sec:xnew} (a = 2.3$\times 10^{-10}$, and b =$-9.6\times 10^{-20}$). Nevertheless, the distribution of this relation is similar to the one derived using \SsfrHa\, with a smaller correlation coefficient as the relation derived in Sec.~\ref{sec:xnew}. The median values of \SsfrSSP\, for different bins of $\mathrm{x_{new}}$ are in good agreement with respect to the best relation derived previously in Sec.~\ref{sec:xnew} (see dashed line) despite the increment in the scatter observed for this relation ($\sigma = [0.39, 0.15]$). Even more, the best fit of these medians is in agreement with the dashed line. 
 
Finally, we plot the \Ssfr-$\mathrm{x_{new}}$ relation following the same procedure as in Sec.~\ref{sec:xnew} (see bottom-right panel in Fig.~\ref{fig:RSsfrSSp}). We note that coefficients of the fit for $\mathrm{x_{new}}$ are different of the ones derived in Sec.\ref{sec:xnew} (a = 2.3$\times 10^{-10}$, and b = -9.6$\times 10^{-20}$). Nevertheless, the distribution of this relation is similar as the one derived using \SsfrHa\, with a smaller correlation coefficient as the relation derived in Sec.~\ref{sec:xnew}. The median values of \SsfrSSP\, for different bins of $\mathrm{x_{new}}$ are in  good agreement with respect to the best relation derived previously in Sec.~\ref{sec:xnew} (see dashed line) despite the increment in the scatter observed for this relation ($\sigma = [0.39, 0.15]$). Furthermore, the best fit the median values derived using \SsfrSSP\, is similar to the one reported using \SsfrHa.

In summary, these results suggest that, independent of the observable we use to determine \Ssfr, we obtain similar trends when we derive the \mbox{\Ssfr - \Sb} relation. Also, the inclusion of an extra quadratic term to describe this relation leads to similar results regardless the \Ssfr\, calibrator. 

\section{Discussion}
\label{sec:Discussion}

In this article, we present the well-known star-forming scaling relations between each of the components of the baryonic mass at kpc scales (i.e., the rSK and rSFMS) for the largest IFU dataset provided by the MaNGA survey ($\sim$ 8000 galaxies). The gas mass density is derived using the optical extinction obtained from the Balmer decrement following Barrera-Ballesteros et al. (in prep.). We further explore different scaling relations between star-formation and the baryonic mass, including a quadratic polynomial of the baryonic mass. We also test these scaling relations using an independent measurement of the \Ssfr\, derived from the SSP analysis.   

The main result of this work is that \Sb provides a better correlation with the \Ssfr\, than using only each of its two components (\Sstar\, and \SmolAv). Even more, when adopting a quadratic polynomial form of \Sb, the log-log relation presents a one-to-one slope (i.e., no power is required to match both quantities). We find similar results when using an independent observable to estimate the star formation rate density. This analysis highlights the necessity to consider the full baryonic content, and not only separate therms of \Sb to properly describe the star formation at kpc scales.  Such a non-linear empirical relations have been pointed out as evidence of the importance of the impact of existing stars in the regulation of SFR \citep[e.g., ][]{Zaragoza-Cardiel_2019}. In this self-regulated model of star-formation, the hydrostatic pressure of the disk galaxy is balanced by the momentum flux injected to the ISM from supernovae explosions \citep[e.g.,][]{Cox_1981, Silk_1997, Ostriker_2010}. From our main result, we argue that a second degree polynomial of \Sb provides a better description of \Ssfr\, since it includes both the contribution of the amount of gas required to form new stars as well as the non-linear terms that describe the impact of the hydrostatic pressure of the disk (see Sec.~\ref{sec:ExtSchmidt}). 

Along this article we measure the scatter of the different scaling relations derived for star-formation regions at kpc scales. For individual components of the baryonic mass we find that the rSFMS yields the smallest scatter in agreement with the scatter derived for a smaller sample of same  MaNGA galaxies \citep[standard deviation, $\sim$ 0.27 dex,][]{Cano-Diaz_2019}. On the other hand, the relation derived with \Sb have slightly smaller scatter with stronger correlation coefficients. These results indicate that when using this specific combination of the baryonic components, the main driver to derive \Ssfr\, in star-forming regions is the stellar mass density. This has also been found in recent studies that classify the strength of correlations among star-formation and other observables at kpc scales. Their results suggest that \Sstar\, appears to be the observable that better correlates with \Ssfr\, \citep{Dey_2019, Bluck_2020}. On the contrary, other explorations, using estimations of the molecular gas based on CO observations of a few tens of objects suggest that the strongest and tighter correlation is found with \Smol \citep{Lin_2019, Ellison_2020}.

Overall, molecular gas is essential to form new stars at kpc scales however the regulation of \Ssfr\, strongly depends on the amount of baryonic matter. In other words, we suggest that locally the gravitational potential is the main regulator of the star formation rate. In Barrera-Ballesteros et al. (submitted) we explore the explicit relation between the star formation and the mid-plane pressure derived from direct estimation of the molecular gas, as well as its interpretation in the context of self-regulation.

\subsection{Other non-linear relations: revisiting the extended Schmidt law}
\label{sec:ExtSchmidt}
%
\begin{figure}
\includegraphics[width=\linewidth]{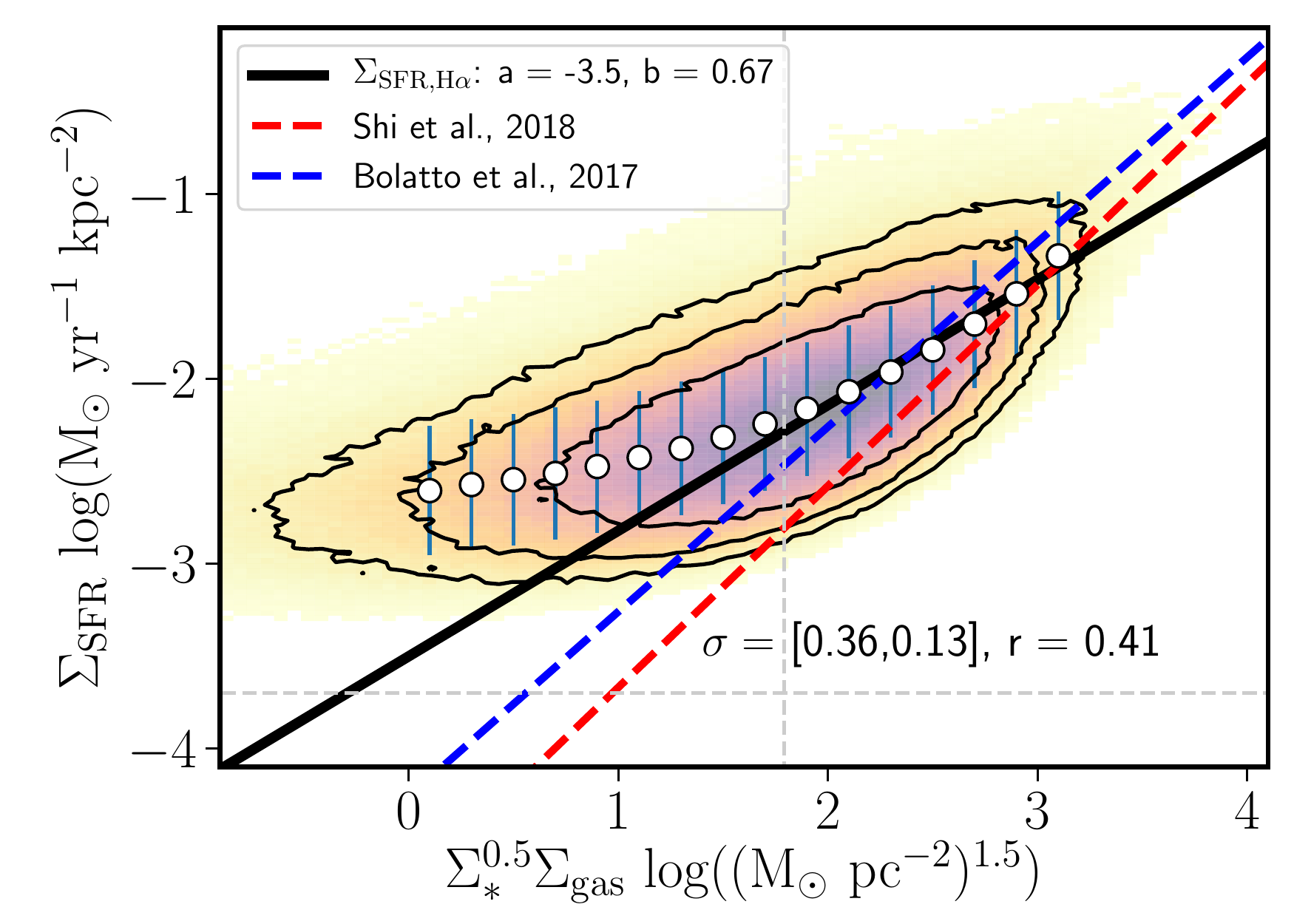}
\caption{The extended star-forming law at kpc scales for the MaNGA sample. The x-axis is defined by \cite{Shi_2011}. The distribution of the selected spaxels (i.e., values larger than the vertical dashed gray line) -- and their median values of \Ssfr\, (white circles) -- follows a similar relation as the one derived by \cite{Shi_2018} (red-dashed line). Although our sample follows the trend of the proposed extended scaling relation by \cite{Shi_2018}, its scatter is similar as the one derived for the rSFMS (see right panel of Fig.~\ref{fig:sKS_SFMS}).}  
\label{fig:Shi_2018}    
\end{figure}

As indicated before, in recent years there have been different studies exploring the relation of a combination of the components of the baryonic mass with the star formation \citep[e.g., ][]{Westfall_2014, Dib_2017, Roychowdhury_2017, Bolatto_2017, delosReyes_2019, Sun_2020}. In particular, \cite{Shi_2011,Shi_2018} explored the so-called extended Schmidt law, which correlates the star-formation surface density with the product of the stellar and gas mass surface density. \cite{Shi_2011} suggest that \Ssfr\, $\sim$ \mbox{$\Sstar^{0.5}\Sgas$} provides a better relation than the Schmidt law. In other words, the scatter of this extended Schmidt law is reduced in comparison to the \Ssfr-\Sgas\, relation. Even more, in \cite{Shi_2018}, they showed that the best relation is slightly super-linear (\mbox{$\Ssfr = 10^{-4.76} (\Sstar^{0.5}\Sgas)^{1.09}$}). In Fig.~\ref{fig:Shi_2018} we explore this extended star-formation law using the current MaNGA data. The slope of the best fit of the median values ($\sim$ 0.67, black solid line) is smaller than one reported by \citep[red dashed line, ][]{Shi_2018}. The slope derived by \cite{Shi_2018} is similar to the one presented by \cite{Bolatto_2017} using the EDGE-CALIFA dataset (blue dashed line). Similarly to the rKS and  rSFMS relations (see Fig.~\ref{fig:sKS_SFMS}), to perform the best fit we use the median values of \Ssfr\, larger than a threshold in the x-axis (\mbox{$10^{1.79}~\mathrm{(M_{\odot}\,\,pc^{-2})^{1.5}}$}, dashed vertical gray line). This threshold considers the limits we use for \SmolAv and \Sstar. The scatter of this relation is smaller than the one we derive for the rSK (standard deviation of 0.36 dex, and 0.45 dex, respectively). However, this scatter has larger dispersion in comparison to the one derived from the rSFMS (0.31 dex). Even more, its correlation coefficient is significantly smaller than the one derived for the rSFMS ($r = 0.41$ vs 0.72). We also note that even though the slope from the best fit is sub-linear, the \Ssfr\,  derived for large  values of \mbox{$\Sgas\Sstar^{0.5}$} is in agreement  with the relations derived in the literature suggesting that for regions with intense star formation, \Sgas\, and \Sstar\, plays an important role describing \Ssfr. In further studies we explore the explicit relation between the \Ssfr\, and the baryonic mass in the context of the self-regulation of star formation in order to quantify the role of the mid-plane pressure in shaping the star-formation rate at kpc scales (Barrera-Ballesteros et al., submitted).   

Recently, \cite{Lin_2019} explored the functional form of the extended KS law (\mbox{$\Sgas\Sstar^{\beta}$}). Using a homogeneous dataset, they found that the exponential that yields the smallest scatter in this relation is $\beta \sim -0.30$. Even when using this exponential the scatter in comparison to the rSK or rSFMS is not reduced as expected from \cite{Shi_2011,Shi_2018}. As these authors, we do not find a strong reduction of the scatter for the star-formation when using the functional form described by \cite{Shi_2018}. From our analysis, we conclude that although a relation such as the extended Schmidt law -- which is derived in the context of self-regulation of star formation -- is necessary to describe the \Ssfr\, it may also needed to include other contributions of the baryonic mass such as the second degree polynomial relation presented in Sec.\ref{sec:xnew}.  

\section{Summary and Conclusions}
\label{sec:Conclusion}

Using a sample of more than 1.1$\times$10$^6$ spatial elements ($\sim$ 3$\times$10$^5$ independent regions) of kpc size located in 2640 galaxies drawn from the MaNGA survey -- the largest IFU survey up to date --, we present a scaling relation between the star formation rate surface density (\Ssfr) and the baryonic mass surface density (\Sb = \Sgas + \Sstar). \Sgas is obtained by using as proxy the optical extinction. Our results can be summarized as follows:

\begin{itemize}

    \item We reproduce the well-known star-forming scaling relations for individual components of \Sb: the resolved Schmidt-Kennicutt law (rSK) and the resolved star-formation main sequence (rSFMS). By measuring their scatter and  correlation factors, we find that the rSFMS yields the tighter and stronger relation with respect to \Ssfr. 

    \item We derive a scaling relation between the \Ssfr, \Sb and a second degree polynomial of \Sb. These relations show a strong correlation and a smaller scatter than those derived from individual components of \Sb. In particular, the second one naturally yields a one-to-one relation. We find similar trends using two independent indicators of \Ssfr: the \ha\, emission line luminosity and a stellar decomposition using single-stellar population fitting of the stellar continuum. 
    
    \item We contrast these new relations with other empirical star-forming scaling relations such as the extended Schmidt law proposed by \cite{Shi_2011, Shi_2018}. We find that the \Ssfr-\Sb relation yields a stronger correlation and has a smaller scatter in comparison to the extended Schmidt law.   
\end{itemize}

We conclude that these star-forming scaling relations quantify the strong impact of the baryonic mass as a whole in the conditions of formation of newly born stars at kpc-scales. Even more, besides the evident role that \Sgas has in the formation of stars, these relations suggest that the local gravitational potential -- measured from the total baryonic mass density --  plays a significant role in shaping the star-formation rate. These results also favor the scenario where star formation is self-regulated at kpc scales. In future studies we explicit study the relation between the hydrostatic pressure of the disk and the star formation rate density at kpc scales.

\acknowledgments
We thank the referee for their comments/suggestions that improves this article. J.B-B and SFS acknowledge support from the grants IA100420 and IN100519 (DGAPA-PAPIIT ,UNAM) and funding from the CONACYT grants CF19-39578, CB-285080, and FC-2016-01-1916. M.B. acknowledges support from the FONDECYT regular grant 1170618. R.A.R thanks partial financial support from Conselho Nacional de Desenvolvimento Cient\'ifico e Tecnol\'ogico (202582/2018-3 and 302280/2019- 7) and Funda\c c\~ao de Amparo \`a pesquisa do Estado do Rio Grande do Sul (17/2551-0001144-9 and 16/2551-0000251-7). P.B.T. acknowledges support from FONDECYT 1200703. Funding for the Sloan Digital Sky Survey IV has been provided by the Alfred P. Sloan Foundation, the U.S. Department of Energy Office of Science, and the Participating Institutions. SDSS acknowledges support and resources from the Center for High-Performance Computing at the University of Utah. The SDSS web site is www.sdss.org.

SDSS is managed by the Astrophysical Research Consortium for the Participating Institutions of the SDSS Collaboration including the Brazilian Participation Group, the Carnegie Institution for Science, Carnegie Mellon University, the Chilean Participation Group, the French Participation Group, Harvard-Smithsonian Center for Astrophysics, Instituto de Astrofísica de Canarias, The Johns Hopkins University, Kavli Institute for the Physics and Mathematics of the Universe (IPMU) / University of Tokyo, the Korean Participation Group, Lawrence Berkeley National Laboratory, Leibniz Institut für Astrophysik Potsdam (AIP), Max-Planck-Institut für Astronomie (MPIA Heidelberg), Max-Planck-Institut für Astrophysik (MPA Garching), Max-Planck-Institut für Extraterrestrische Physik (MPE), National Astronomical Observatories of China, New Mexico State University, New York University, University of Notre Dame, Observatório Nacional / MCTI, The Ohio State University, Pennsylvania State University, Shanghai Astronomical Observatory, United Kingdom Participation Group, Universidad Nacional Autónoma de México, University of Arizona, University of Colorado Boulder, University of Oxford, University of Portsmouth, University of Utah, University of Virginia, University of Washington, University of Wisconsin, Vanderbilt University, and Yale University.

\appendix
\section{\Ssfr\, estimations}
\label{app:SFRs}

\begin{figure}
\includegraphics[width=\linewidth]{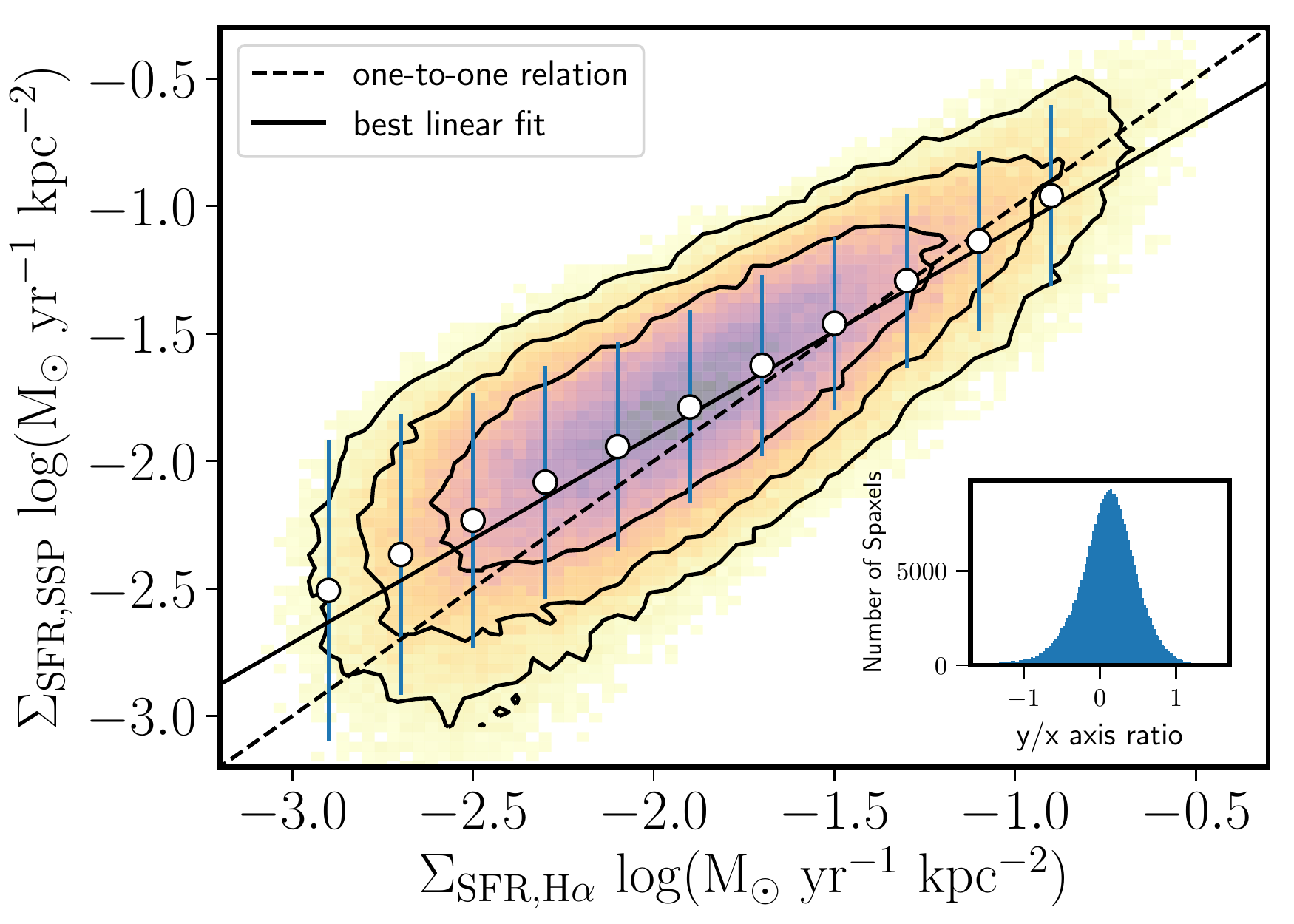}
\caption{A comparison between the star formation derived from the SSP analysis, \SsfrSSP, against the one derived using the \ha\, luminosity \SsfrHa. As in previous plots, the contours enclose 90\%, 80\%, and 60\% of the distribution. The white circles show the median \SsfrSSP\, for bins of  \SsfrHa. The solid lines shows the best fit of these bins whereas the dashed lines represents the one-to-one relation. Both estimations of \Ssfr\, are similar to each other.}  
\label{fig:SFRs}   
\end{figure}

In Sec.~\ref{sec:SSPs} we use the \Ssfr\, derived from the SSP to estimate the star-forming scaling relations derived using the \ha\, luminosity as proxy of \Ssfr. In this appendix we compare these two independent estimations of this quantity. In Fig.~\ref{fig:SFRs} we compare the \Ssfr\, derived using SSPs against the one derived \ha\, luminosity for the sample of spaxels selected in  Sec.\ref{sec:SSPs}.  

\bibliographystyle{mnras}
\bibliography{main}

\end{document}